\documentclass[iop]{emulateapj}

\usepackage{natbib}
\usepackage{ulem}
\usepackage{epsfig}
\usepackage{amsmath}

\shorttitle{AGN Driven Metal Outflows}
\shortauthors{Kirkpatrick, McNamara \& Cavagnolo}
\slugcomment{Accepted for publication in ApJ Letters}

\begin{document}

\title{Anisotropic Metal-enriched Outflows Driven by AGN in Clusters of Galaxies}

\author{C.~C. Kirkpatrick\altaffilmark{1}, 
              B.~R. McNamara\altaffilmark{1,2,3}
              \& K.~W. Cavagnolo\altaffilmark{1,4}}

\affil{
\altaffilmark{1}Department of Physics \& Astronomy, University of Waterloo, 200 University Ave. W., Waterloo, ON N2L 3G1, Canada \\
\altaffilmark{2}Perimeter Institute for Theoretical Physics, 31 Caroline St. N., Waterloo, ON N2L 2Y5, Canada \\
\altaffilmark{3}Harvard-Smithsonian Center for Astrophysics, 60 Garden St., Cambridge, MA 02138 \\
\altaffilmark{4}UNS, CNRS UMR 6202 Cassiop\'{e}e, Observatoire de la C\^{o}te d'Azur, Nice, France
}


\begin{abstract}
We present an analysis of the spatial distribution of metal-rich gas in ten galaxy clusters using deep observations from the {\it Chandra X-ray Observatory}.  The brightest cluster galaxies have experienced recent AGN activity in the forms of bright radio emission, cavities, and shock fronts embedded in the hot atmospheres.  The heavy elements are distributed anisotropically and are aligned with the large-scale radio and cavity axes.  They are apparently being transported from the halo of the brightest cluster galaxy into the intracluster medium along large-scale outflows driven by the radio jets.  The radial ranges of the metal-enriched outflows are found to scale with jet power as $R_{\rm Fe} \propto P_{\rm jet}^{0.42}$, with a scatter of only 0.5 dex.  The heavy elements are transported beyond the extent of the inner cavities in all clusters, suggesting this is a long lasting effect sustained over multiple generations of outbursts.  Black holes in BCGs will likely have difficulty ejecting metal enriched gas beyond 1 Mpc unless their masses substantially exceed $10^9$ M$_\sun$.
\end{abstract}
 

\section{Introduction}

A hot, diffuse plasma composed primarily of hydrogen and helium and enriched in heavy elements fills the space in and between the galaxies in rich clusters.  Chemical enrichment of the intracluster medium (ICM) is thought to have occurred early, primarily by type II supernova explosions.  The supernova ejecta raised the heavy element abundance of the ICM to an average of approximately 1/3 of the solar value \citep{mus96,mus97}.  Exceptions are found near brightest cluster galaxies (BCGs) located at the centers of cool core (cooling flow) clusters.  The  heavy element abundances there often rise to values approaching the mean metallicity of the sun \citep{all98,dup00,deg01}.  Furthermore, the iron abundance in cool cores is usually enhanced with respect to the $\alpha$-elements.  Because the $\alpha$ elements  are produced primarily by core collapse supernovae, while iron is produced primarily by type Ia supernovae, the gas in cool cores has probably been enriched relative to the ICM by the stars associated with the brightest cluster galaxies \citep{deg04,tam04}.  

Several studies have shown that the radial distributions of type Ia ejecta extend to larger radii than the stellar light profiles of BCGs \citep{reb05,reb06,dav08,ras08}.  Assuming the ejecta was produced by the stars,  the profiles imply that the ejecta is being displaced from the galaxies by some mechanism.  Turbulence induced by cluster mergers or outflows from active galactic nuclei (AGN) are likely possibilities \citep{sha09}.  BCGs often host radio AGN that are interacting with the hot, metal-rich gas by forming cavity systems and shock fronts \citep[see][for review]{bmc07}.  The cavity systems in the objects discussed here have displaced between $10^{10}\sun$ and $10^{12}\sun$ of gas.  The dynamical forces associated with them may also be able to drive the metal-enriched gas out of the BCG and into the ICM in outflows oriented along the radio jet \citep{gop01,gop03,bru02,omm04,mol07,roe07,pop10}.

Recent studies of some clusters have identified gas with unusually high metallicity beyond the cores and along the radio lobes \citep{sim08,sim09,cck09b,osul11}.  The gas along the radio lobes is, in some cases, enhanced in metallicity by up to 0.2 dex relative to the surrounding gas.  These metal-rich inclusions indicate that the cool, metal-rich material originating from the BCG is mixing with lower metallicity gas lying at larger radii.  It is still unclear how general this phenomenon is, and how these outflows scale with the properties of the host AGN.  A systematic analysis of a large sample of X-ray clusters with a range in AGN jet power is needed to definitively explore this phenomenon.

Here we show that AGN are driving metals out of BCGs to very large distances.  We have performed our analysis using deep, high resolution X-ray images of ten clusters drawn from the {\it Chandra X-ray Observatory} archive.  We report on the discovery of a significant trend between jet power and the radial extent of uplifted, metal-enriched gas, and we briefly discuss its consequences.  Throughout this letter we assume a $\Lambda$CDM cosmology with H$_0 = 70$ km s$^{-1}$ Mpc$^{-1}$, $\Omega_{\rm M} = 0.3$, and $\Omega_{\Lambda} = 0.7$.  All uncertainties are quoted at the 67\% confidence level.

\begin{figure*}[t]
\begin{center}
\begin{minipage}{0.326\linewidth}
\includegraphics*[width=\textwidth, trim=0mm 0mm 0mm 0mm, clip]{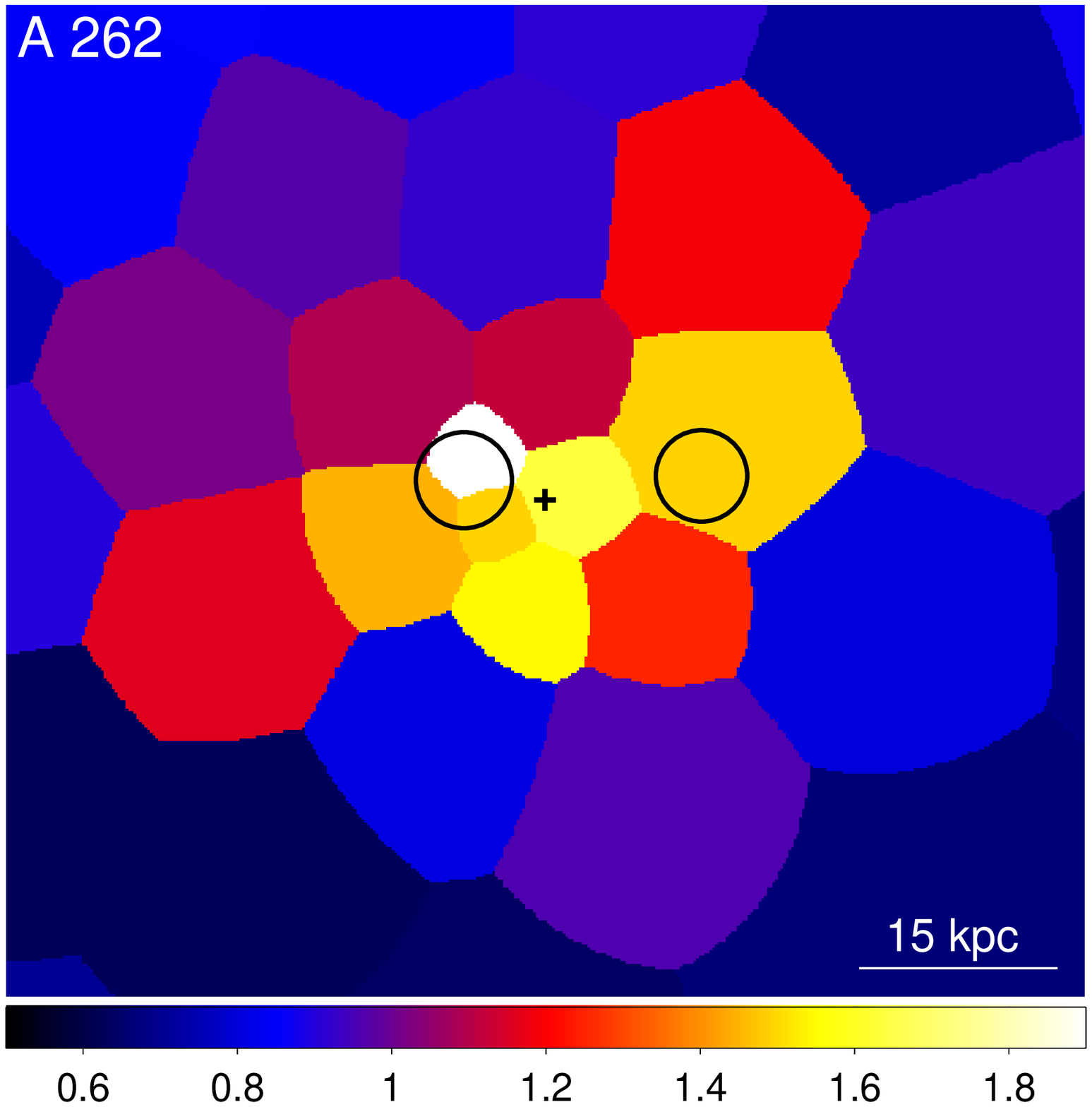}
\end{minipage}
\begin{minipage}{0.326\linewidth}
\includegraphics*[width=\textwidth, trim=0mm 0mm 0mm 0mm, clip]{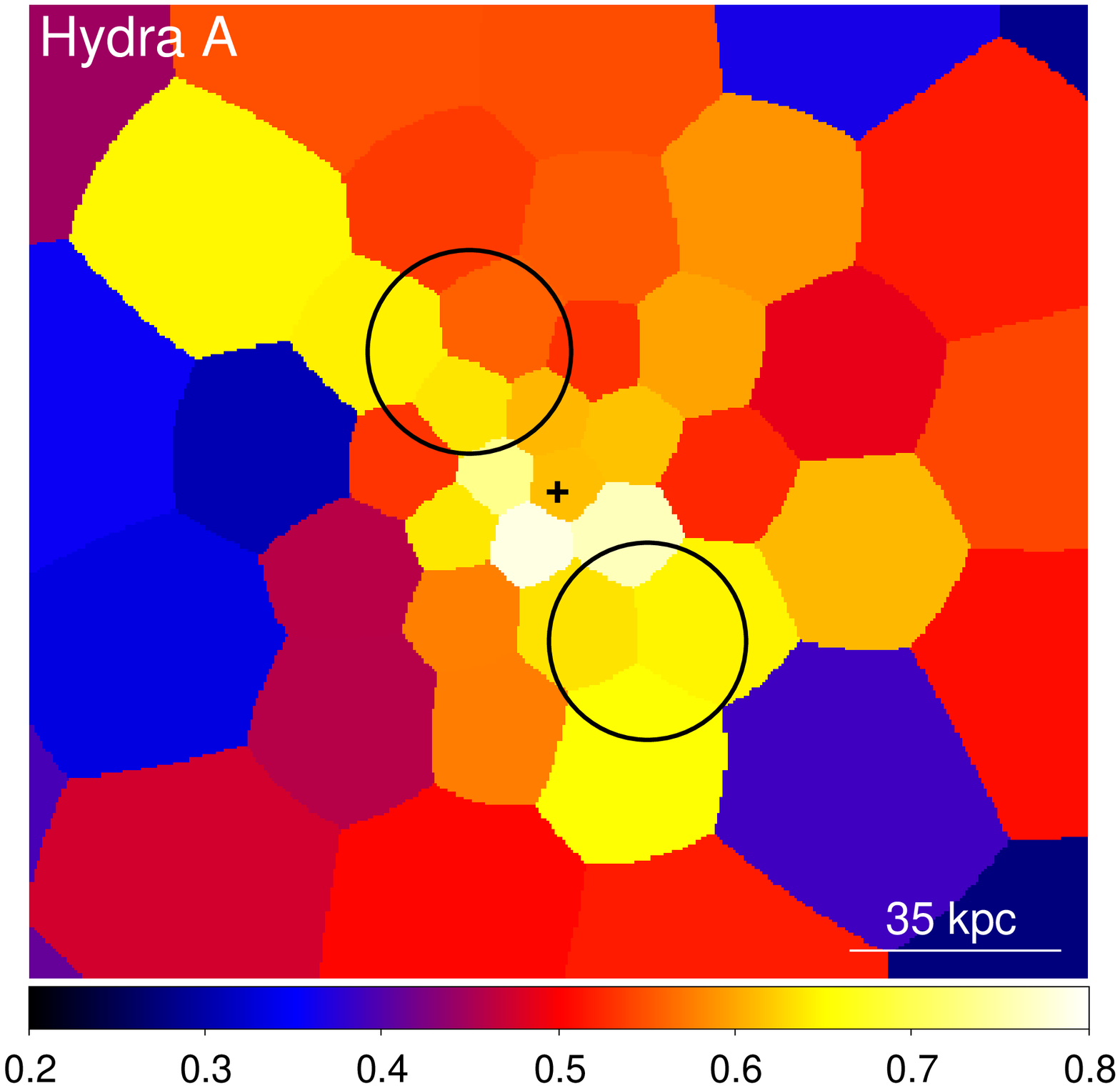}
\end{minipage}
\begin{minipage}{0.326\linewidth}
\includegraphics*[width=\textwidth, trim=0mm 0mm 0mm 0mm, clip]{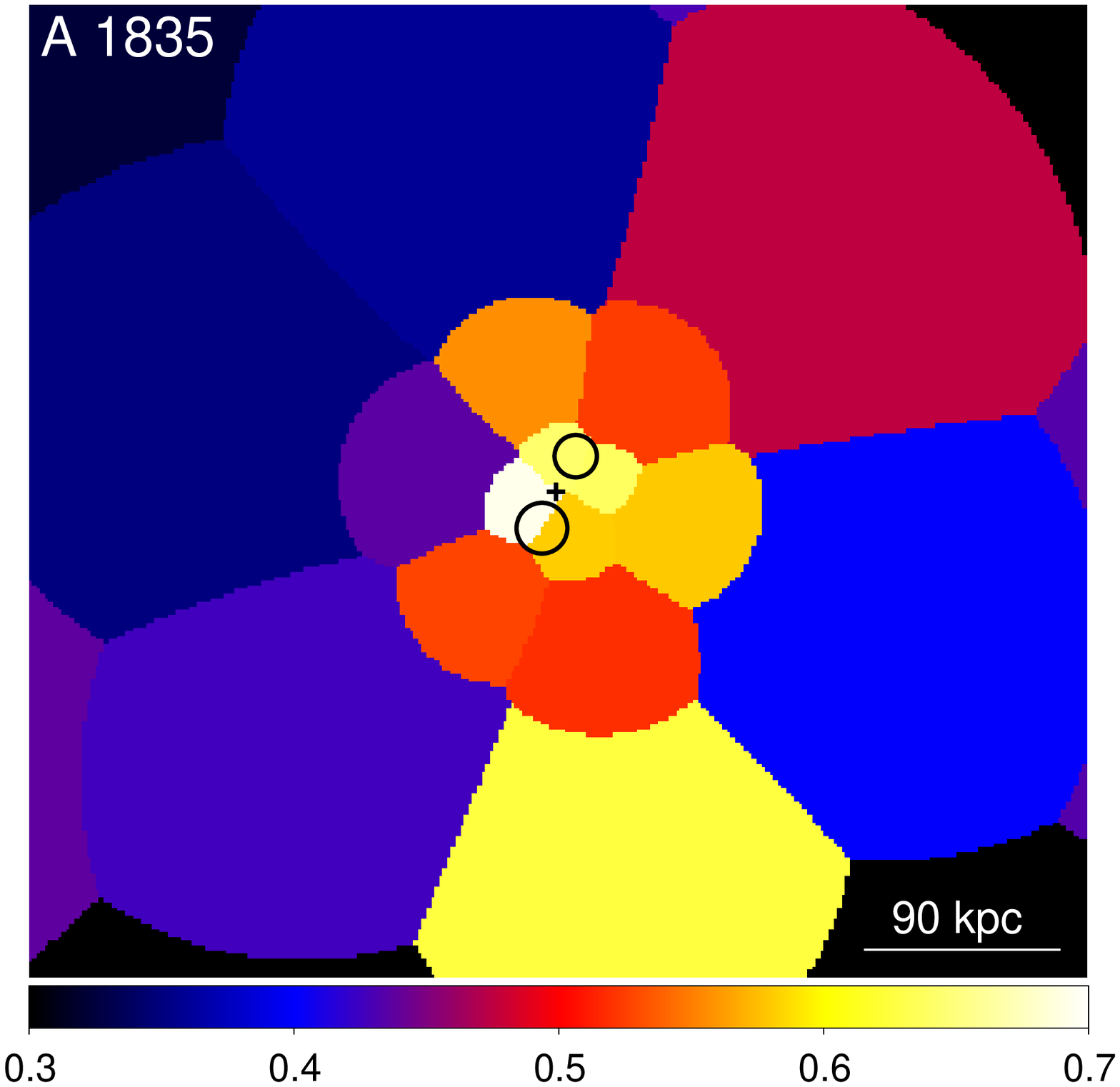}
\end{minipage}
\caption{The three panels show the metallicity maps for clusters Abell 262, Hydra A, and Abell 1835, respectively.  In all three cases, high metallicity gas (shown in yellow) extends outward along the same axis of the cavity system (approximate size and location indicated by black circles).  The black crosses indicate the approximate cluster centers.}
\end{center}
\end{figure*}

\section{Cluster Sample}

We examine 10 clusters selected from the \citet{bir04}, \citet{raf06} and \citet{die08} samples of clusters with prominent cavity systems.  The ten clusters span a broad range of jet (cavity) power.  The sample includes clusters having Chandra images that are deep enough to generate metallicity maps with a resolution of at least 25 bins with a minimum signal-to-noise of 100 per bin.  Our goal is to make high precision measurements of the projected metallicity distribution while maintaining a spatial resolution that optimizes the measurement of the spatial distribution of the high metallicity gas for comparison to the radio sources.  The clusters that meet these criteria are A133, A262, Perseus, MS0735.6+7421, Hydra A, M87, A1835, A2029, A2199, and A2597.  Each cluster observation was processed using version 4.1.2 of CIAO and the calibration database.  Background flares were excluded using standard filtering techniques and point sources were identified in the combined image of each cluster using {\it wavedetect}, and removed.  Blank-sky background files were normalized to the source count rate in the 9.5-12 keV band.

\section{Analysis}

\subsection{Metallicity Maps and Profiles}

We present example metallicity maps for three clusters (Abell 262, Hydra A, Abell 1835) in Figure 1.  The combined exposures for each cluster were binned using a weighted Voronoi tessellation algorithm \citep{cap03,die06}.  With approximately 10000 net counts or greater per bin, spectra were extracted and weighted response matrices were created using standard CIAO routines.  A single temperature plasma model with absorption (WABS$\times$MEKAL) was fit over the energy range 0.5-7 keV.  Temperature, abundance, and normalization were allowed to vary with the column density frozen at the values quoted by \citet{dic90}.  The internal ratio between metals were set to the solar photospheric values of \citet{gre98}.  Allowing the $\alpha$-elements to vary independently does not show any significant effect on the outcome of the iron abundance measurement, therefore we have chosen to use the simple model.  The spectra are reasonably well fit by the single temperature model, with the exception of the central region of A262.  This region is better fit with a multi-temperature model which finds a higher central metallicity value \citep[see][]{cck09a}.  However, this does not affect the interpretation of our results as the central regions of the clusters are not important for this analysis.

Metallicity profiles have been measured for each cluster.  Spectra were extracted from regions occupied by radio emission and/or cavities in semi-annular shaped bins.  Each bin size was determined by requiring a minimum signal-to-nose of 140, which achieves uncertainties of approximately 0.05 Z$_\sun$.  A high signal to noise ratio is required to find deviations between the profiles along and orthogonal to the radio cavity system.  These orthogonal profiles, which we consider to represent the undisturbed region of the cluster, are extracted using the same bin width as the radio cavity system profile.  The S/N for these regions are usually greater than 140 due to larger angular sizes.  All profiles are fit using the same model as the 2D analysis.

\subsection{P$_{\rm jet}$-R$_{\rm Fe}$ Scaling Relation}

The radial profiles for A262, Hydra A, and A1835 are presented in Figure 2.  The red squares represent the metallicity of the radio cavity system and the blue triangles represent the undisturbed metallicity profile.  The dotted vertical line indicates the maximum radius at which a significant enhancement in metallicity has been detected.  We refer to this as the iron radius ($R_{\rm Fe}$).  This radius is concisely defined as the location of the radial bin furthest from the cluster center where the one sigma error bars of the profiles along and orthogonal to the radio axis do not overlap.  Regions interior to the iron radius show a systematic enhancement in iron abundance over the undisturbed regions.  For regions beyond the iron radius, the abundance profiles are indistinguishable.  From top to bottom the three panels are arranged in order of increasing jet power and increasing iron radius.  This shows that higher power AGN are able to launch metal enriched gas to greater distances. 

\begin{figure}[t]
\epsscale{1.27}
\plotone{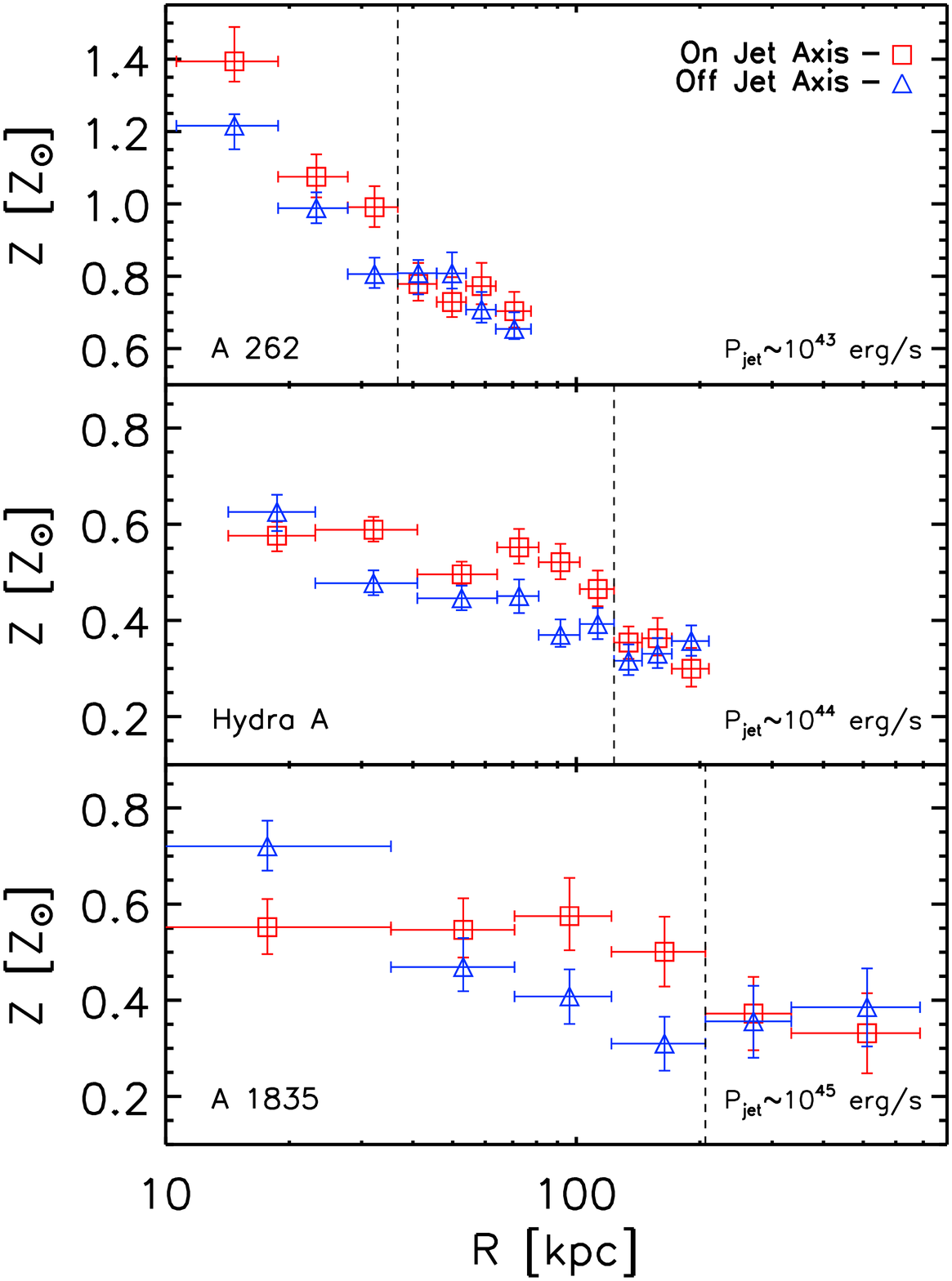}
\caption{Each panel represents metallicity profiles along and orthogonal to the radio axis for each cluster.  The red square points represent the metallicity along the axis of the radio cavity system, and the blue triangular points represent the undisturbed region of each cluster.  The dashed vertical line represents the iron radius.}
\end{figure}

In Figure 3, the iron radius is plotted against jet power derived using X-ray cavity data taken from \citet{raf06} for all ten clusters in the sample.  The powers quoted there were derived from the total energy of the cavities, $4pV$, and their buoyant rise time.  In systems with multiple generations of cavities, the average power is used.  Note that the iron radius was determined using the radial metallicity profiles only and not using the metallicity maps.  A trend between jet power and iron radius is evident over three decades in jet power.  The low powered AGN outbursts $P_{\rm jet} \simeq 10^{43}\rm~erg~s^{-1}$ drive out material on a scale of a few tens of kiloparsecs.  For exceptionally large outbursts with jet power exceeding $10^{46}\rm~erg~s^{-1}$, such as MS0735, metals from the core are being launched hundreds of kiloparsecs into the ICM.

To quantify this trend we fit a linear function to the logarithms of jet power and iron radius.  Performing a least squares regression, the best fit plotted in Figure 3 takes the form
\begin{equation}
R_{\rm Fe} = 58 \times P_{\rm jet}^{0.42} ~(\rm kpc),
\end{equation}
where jet power is in units of $10^{44}$ ergs s$^{-1}$.  The RMS scatter about the fit is approximately 0.5 dex.  The correlation is strong and shows a fairly small scatter, although with only ten objects the true scatter about the mean is difficult to evaluate.  The scatter will likely increase with the inclusion of additional clusters, which will be addressed in a future paper.  

\begin{figure}[t]
\epsscale{1.27}
\plotone{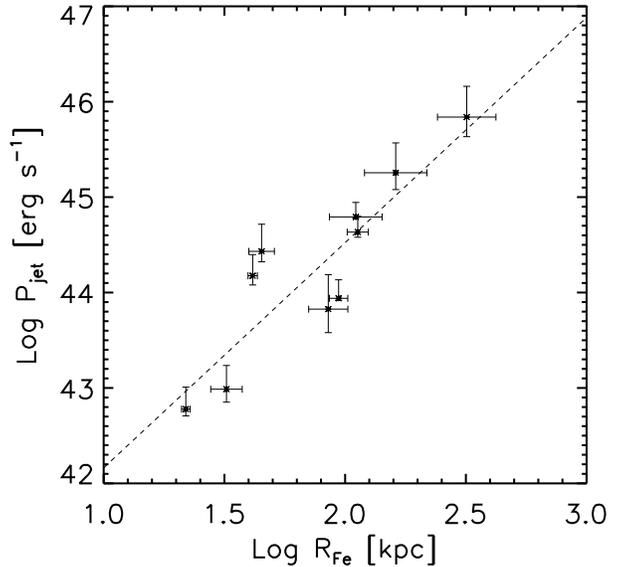}
\caption{Jet power versus iron radius.  The dashed line is the best fit to the data.}
\end{figure}

A spurious correlation between iron radius and jet power, which depends on the volume of the cavities, could arise due to the dependence of distance on both the linear diameter of the cavity systems and the iron radius.  This does not appear to be the case.  The clusters we considered here all have been exposed deeply with Chandra.  We could have easily detected cool, metal-rich gas on all spatial scales of interest here, and at random angles with respect to the cavities and radio sources. We did not.  Hydra A, for example, has a complex cavity system that was created by at least three separate outbursts or a continuous outburst that has persisted for several hundred million years \citep{wis07}.  The cavity systems considered separately have similar jet powers, but their centers lie at very different radii between 25 kpc and 250 kpc.  Nevertheless, Hydra A does not deviate significantly from the trend in Figure 3, implying that the measured iron radius is not a simple function of cavity size alone.  In Figure 5 we have plotted the inner cavity position for all clusters in our sample versus the iron radius.  In all cases the iron radius lies beyond the inner cavities.  This shows that the metallicity outflows are not simply tracing the current generation of AGN activity, but are maintained over multiple generations.  The iron radius may provide a reliable indicator of average jet power in a cluster where cavity power measurements are ambiguous.

There are uncorrected systematic uncertainties in Figure 3.  We have made no attempt to measure the additional power associated with shock fronts and faint ghost cavity systems, which would be difficult to do for the entire sample.  The total jet power for objects such as MS0735 and Hydra A, which have detected shock fronts at high significance, are under reported here by roughly a factor of two (see McNamara \& Nulsen 2007 for a discussion).  Including energy from the shocks effects the slope of the fit by $\sim 20\%$.  We chose not to boost the power of these systems to avoid biasing them with respect to other systems.  Thus, only cavity power is given here for consistency.  In addition, the iron radii given here are projected values. Their true values may be underestimated in extreme cases by roughly a factor of two, but typically by a few tens of percent.  Errors of this size will not change our basic result.  We intend to explore these issues in a future paper.

\subsection{Alignment Between Metal Enhancements and the Radio Orientation}

We find that the high metallicity gas outflow regions are spatially aligned with the radio and X-ray cavity systems.  Using the metallicity maps, we have independently measured the angle on the sky with respect to the center of the cluster of the high metallicity gas outside of the cluster core.  We have also measured the mean position angle of the cavity system.  The position angle was measured using unsharp masks of each cluster in order to estimate the locations of the centers of the cavities.  Finally, the angle of the extended radio emission was determined for those systems with resolved radio imagery.  Using 330 MHz images, the radio angle was determined by bisecting the radio emission on either side of the nucleus of the BCG.  We have plotted metal angle versus cavity center (blue squares) and radio angle (red diamonds) in Figure 4.  Both sets of points are consistent with the dashed line of equality.  In clusters with calculated radio angles, all three quantities are in excellent agreement with one another.

\begin{figure}[t]
\epsscale{1.25}
\plotone{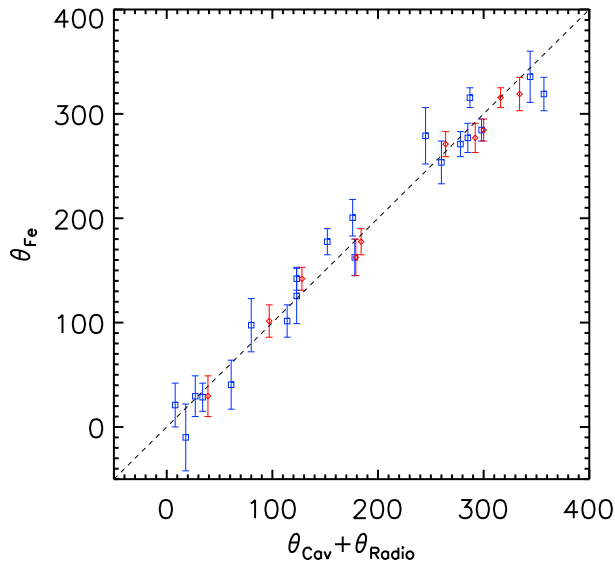}
\caption{Angle of metal enhanced gas versus cavity angle and radio emission angle.  The blue squares represent angle on the sky with respect to the center of the cavity.  Where high resolution data is available, the radio angle (330 MHz radio emission) is represented by red diamonds.  Blue and red points located at the same $\theta_{\rm Fe}$ highlight the difference in cavity and jet angle in the same cluster.  The dashed line represents the line of equality.}
\end{figure}

\subsection{Comparison to Independent Metallicity Measurements and to Simulations}

Independent reports of metal enriched outflows are consistent  with the trend shown in Figure 3. For example, the radial abundance effect is seen in a relatively low S/N metallicity map of the poor cluster AWM 4 \citep{osul10,osul11}.  This study found metal-enriched gas extending along the radio jets from the nucleus of the BCG to a radius of 35 kpc.  With a jet power of approximately $2 \times 10^{43}$ erg s$^{-1}$, our scaling relation predicts an iron radius of 25 kpc.  This result lies within one standard deviation of our relationship.

\begin{figure}[t]
\epsscale{1.25}
\plotone{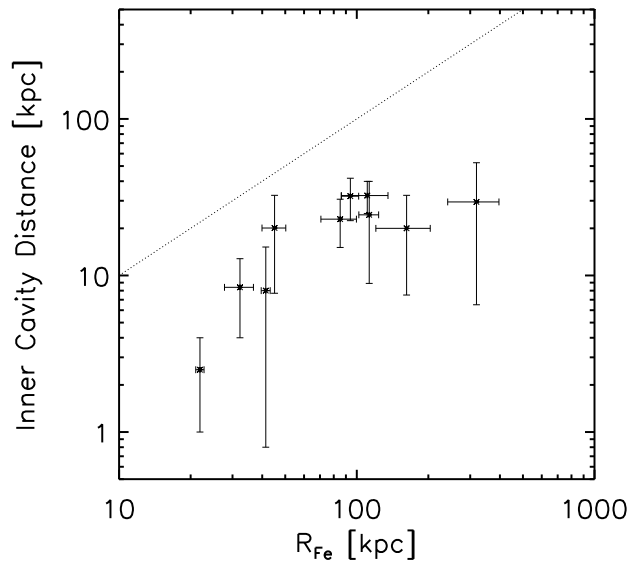}
\caption{The radius of the innermost cavity of each cluster is plotted against the iron radius.  The points represent the cavity center and the vertical error bars indicate the radius of the cavity.  The dotted line of equality is also plotted.}
\end{figure}

Recent simulations of AGN outflows have shown that radio jets are expected to advect ambient material and launch it into the ICM \citep[eg.,][]{pop10}.  Using a 3D AMR hydrodynamical code,  \citet{gas10} found in their simulations that jets are able to transport metal-rich gas from the cores of clusters into the ICM.  Their simulated jet power of approximately $5.7 \times 10^{44}$ erg s$^{-1}$ lifted gas 200 kpc from the center before it mixed with the surrounding metal-poor gas.  The simulation over-predicts the iron radius by a factor of 1.6 compared to our scaling relation, which represents about a one sigma deviation from our relationship.

Using a series of three dimensional hydrodynamical jet simulations centered in a realistic, non-static atmosphere,  \citet{mor10} have evaluated the spatial impact of radio jets on the hot ICM as a function of jet power.  They refer to this as the radial  ``sphere of influence" of the jet, which we compare to our iron radius.   They find that the sphere of influence scales with jet power as $R \propto L_{\rm jet}^{1/3}$.  This dependence on jet power is somewhat shallower than ours.  However, the rough agreement between our observed scaling relation and their simulations is encouraging.

\section{Summary \& Discussion}

We have shown that metal-rich gas is being transported outward by AGN in all ten clusters in our sample.  In each case, the metallicity enhancements found in the maps are closely aligned with the radio and cavity systems.  We find a significant trend between jet power and iron radius, with a power-law scaling as $R_{\rm Fe} \propto P_{\rm jet}^{0.42}$.  This scaling relation shows that AGN play an important role in dispersing metals throughout the cores of clusters.  A comparison between our scaling relation and recent hydrodynamical simulations discussed in the literature shows encouraging agreement.  A detailed comparison between observed and simulated outflows shows strong potential to place interesting constraints on jet physics. 

Simulations of cluster assembly seem to require outflows in order to spread metals throughout the cores of clusters \citep{fab10,bar11}.  Major mergers at early times are efficient at distributing gas throughout the ICM \citep{bur08}, but recent studies show in cool core clusters, sloshing due to mergers alone cannot explain at later times the large radial extent (i.e. $\ga 100$ kpc) of metallicity peaks \citep{roe10}.  A simple extrapolation of our scaling relation to a higher jet power indicates that AGN exceeding $10^{47} ~\rm erg~s^{-1}$ would be required to drive metals beyond 1 Mpc, which is a cosmologically interesting regime.  Mechanical jet powers of this magnitude have not been observed in clusters, although quasars radiating above $10^{47}~\rm erg~s^{-1}$ have been observed in the centers of a few cool core clusters \citep[eg.,][]{rus10}.   Achieving mechanical jet powers of this magnitude requires accretion rates of several tens of solar masses per year onto a nuclear black hole.  At the same time,  mechanically dominant AGN are expected to occur in radiatively inefficient inflows accreting below a few percent of the Eddington rate \citep{nar08}.  Therefore achieving such powerful mechanical outflows would be exceedingly difficult unless the host galaxies harbor black holes with masses significantly in excess of $10^9~\rm M_\odot$, which are expected to be rare.  This may explain why the metallicity enhancements at the centers of clusters extend several hundred kpc beyond the BCG and generally not to 1 Mpc.

\acknowledgments

We would like to thank Paul Nulsen and Mike Wise for helpful comments.  This work was funded in part by {\it Chandra} grant G07-8122X and a generous grant from the Natural Science and Engineering Research Council of Canada.


\begin{thebibliography}{}
\bibitem[Allen \& Fabian(1998)]{all98} Allen, S.~W., \& 
Fabian, A.~C.\ 1998, \mnras, 297, L63
\bibitem[Barai et al.(2011)]{bar11} Barai, P., Martel, H., 
\& Germain, J.\ 2011, \apj, 727, 54
\bibitem[B{\^i}rzan et al.(2004)]{bir04} B{\^i}rzan, L., Rafferty, D.~A., 
McNamara, B.~R., Wise, M.~W., \& Nulsen, P.~E.~J.\ 2004, \apj, 607, 800
\bibitem[Br{\"u}ggen(2002)]{bru02} Br{\"u}ggen, M.\ 2002, 
\apjl, 571, L13
\bibitem[Burns et al.(2008)]{bur08} Burns, J.~O., Hallman, E.~J., Gantner, B., 
Motl, P.~M., \& Norman, M.~L.\ 2008, \apj, 675, 1125
\bibitem[Cappellari \& Copin(2003)]{cap03} Cappellari, M., \& 
Copin, Y.\ 2003, \mnras, 342, 345
\bibitem[David \& Nulsen(2008)]{dav08} David, L.~P., \& 
Nulsen, P.~E.~J.\ 2008, \apj, 689, 837
\bibitem[De Grandi \& Molendi(2001)]{deg01} De Grandi, S., \& 
Molendi, S.\ 2001, \apj, 551, 153
\bibitem[De Grandi et al.(2004)]{deg04} De Grandi, S., Ettori, S., 
Longhetti, M., \& Molendi, S.\ 2004, \aap, 419, 7
\bibitem[Diehl \& Statler(2006)]{die06} Diehl, S., \& 
Statler, T.~S.\ 2006, \mnras, 368, 497
\bibitem[Diehl et al.(2008)]{die08} Diehl, S., Li, H., Fryer, 
C.~L., \& Rafferty, D.\ 2008, \apj, 687, 173
\bibitem[Dickey \& Lockman(1990)]{dic90} Dickey, J.~M., \& 
Lockman, F.~J.\ 1990, \araa, 28, 215
\bibitem[Dupke \& White(2000)]{dup00} Dupke, R.~A., \& 
White, R.~E., III 2000, \apj, 537, 123
\bibitem[Fabjan et al.(2010)]{fab10} Fabjan, D., Borgani, S., Tornatore, L., 
Saro, A., Murante, G., \& Dolag, K.\ 2010, \mnras, 401, 1670
\bibitem[Gaspari et al.(2011)]{gas10} Gaspari, M., Melioli, 
C., Brighenti, F., \& D'Ercole, A.\ 2011, \mnras, 411, 349
\bibitem[Gopal-Krishna \& Wiita(2001)]{gop01} Gopal-Krishna, \& 
Wiita, P.~J.\ 2001, \apjl, 560, L115
\bibitem[Gopal-Krishna \& Wiita(2003)]{gop03} Gopal-Krishna, \& 
Wiita, P.~J.\ 2003, Radio Astronomy at the Fringe, 300, 293
\bibitem[Grevesse \& Sauval(1998)]{gre98} Grevesse, N., \& Sauval, 
A.~J.\ 1998, \ssr, 85, 161
\bibitem[Kirkpatrick et al.(2009a)]{cck09a} Kirkpatrick, C.~C., McNamara, 
B.~R., Rafferty, D.~A., Nulsen, P.~E.~J., B{\^i}rzan, L., Kazemzadeh, F., 
Wise, M.~W., Gitti, M. \& Cavagnolo, K.~W.\ 2009, \apj, 697, 867
\bibitem[Kirkpatrick et al.(2009b)]{cck09b} Kirkpatrick, C.~C., 
Gitti, M., Cavagnolo, K.~W., McNamara, B.~R., David, L.~P., Nulsen, 
P.~E.~J., \& Wise, M.~W.\ 2009, \apjl, 707, L69
\bibitem[Moll et al.(2007)]{mol07} Moll, R., Schindler, S., 
Domainko, W., Kapferer, W., Mair, M., van Kampen, E., Kronberger, T., 
Kimeswenger, S. \& Ruffert, M.\ 2007, \aap, 463, 513
\bibitem[McNamara \& Nulsen(2007)]{bmc07} McNamara, B.~R., \& 
Nulsen, P.~E.~J.\ 2007, \araa, 45, 117
\bibitem[Morsony et al.(2010)]{mor10} Morsony, B.~J., Heinz, 
S., Br{\"u}ggen, M., \& Ruszkowski, M.\ 2010, \mnras, 407, 1277
\bibitem[Mushotzky et al.(1996)]{mus96} Mushotzky, R., Loewenstein, 
M., Arnaud, K.~A., Tamura, T., Fukazawa, Y., Matsushita, K., Kikuchi, K., \& 
Hatsukade, I.\ 1996, \apj, 466, 686
\bibitem[Mushotzky \& Loewenstein(1997)]{mus97} Mushotzky, R.~F., \& 
Loewenstein, M.\ 1997, \apjl, 481, L63
\bibitem[Narayan \& McClintock(2008)]{nar08} Narayan, R., \& 
McClintock, J.~E.\ 2008, New Astron. Rev., 51, 733 
\bibitem[Omma et al.(2004)]{omm04} Omma, H., Binney, J., 
Bryan, G., \& Slyz, A.\ 2004, \mnras, 348, 1105
\bibitem[O'Sullivan et al.(2010)]{osul10} O'Sullivan, E., Giacintucci, 
S., David, L.~P., Vrtilek, J.~M., \& Raychaudhury, S.\ 2010, \mnras, 407, 321
\bibitem[O'Sullivan et al.(2011)]{osul11} O'Sullivan, E., Giacintucci, S., 
David, L.~P., Vrtilek, J.~M., \& Raychaudhury, S.\ 2011, \mnras, 411, 1833
\bibitem[Pope et al.(2010)]{pop10} Pope, E.~C.~D., Babul, A., Pavlovski, G., 
Bower, R.~G., \& Dotter, A.\ 2010, \mnras, 406, 2023
\bibitem[Rafferty et al.(2006)]{raf06} Rafferty, D.~A., 
McNamara, B.~R., Nulsen, P.~E.~J., \& Wise, M.~W.\ 2006, \apj, 652, 216
\bibitem[Rasera et al.(2008)]{ras08} Rasera, Y., Lynch, B., 
Srivastava, K., \& Chandran, B.\ 2008, \apj, 689, 825
\bibitem[Rebusco et al.(2005)]{reb05} Rebusco, P., Churazov, 
E., B{\"o}hringer, H., \& Forman, W.\ 2005, \mnras, 359, 1041
\bibitem[Rebusco et al.(2006)]{reb06} Rebusco, P., Churazov, 
E., B{\"o}hringer, H., \& Forman, W.\ 2006, \mnras, 372, 1840
\bibitem[Roediger et al.(2007)]{roe07} Roediger, E., Br{\"u}ggen, M., Rebusco, 
P., B{\"o}hringer, H., \& Churazov, E.\ 2007, \mnras, 375, 15
\bibitem[Roediger et al.(2010)]{roe10} Roediger, E., Br{\"u}ggen, M., Simionescu, A., 
B{\"o}hringer, H., Churazov, E., \& Forman, W.~R.\ 2010, arXiv:1007.4209
\bibitem[Russell et al.(2010)]{rus10} Russell, H.~R., Fabian, 
A.~C., Sanders, J.~S., Johnstone, R.~M., Blundell, K.~M., Brandt, W.~N., 
\& Crawford, C.~S.\ 2010, \mnras, 402, 1561 
\bibitem[Sharma et al.(2009)]{sha09} Sharma, P., Chandran, 
B.~D.~G., Quataert, E., \& Parrish, I.~J.\ 2009, \apj, 699, 348
\bibitem[Simionescu et al.(2008)]{sim08} Simionescu, A., Werner, N., 
Finoguenov, A., B{\"o}hringer, H., \& Br{\"u}ggen, M.\ 2008, \aap, 482, 97
\bibitem[Simionescu et al.(2009)]{sim09} Simionescu, A., Werner, N., 
B{\"o}hringer, H., Kaastra, J.~S., Finoguenov, A., Br{\"u}ggen, M., \& 
Nulsen, P.~E.~J.\ 2009, \aap, 493, 409
\bibitem[Tamura et al.(2004)]{tam04} Tamura, T., Kaastra, J.~S., den Herder, 
J.~W.~A., Bleeker, J.~A.~M., \& Peterson, J.~R.\ 2004, \aap, 420, 135
\bibitem[Wise et al.(2007)]{wis07} Wise, M.~W., McNamara, B.~R., Nulsen, 
P.~E.~J., Houck, J.~C., \& David, L.~P.\ 2007, \apj, 659, 1153
\end{thebibliography}
\end{document}